\documentclass{article}
\usepackage{graphicx} 
\usepackage{amsmath}
\usepackage{caption}
\usepackage{url}
\usepackage{subcaption}
\usepackage{tikz}
\usetikzlibrary{shapes.geometric, arrows}

\title{Magnetic Compressors for MeV-UED}
\author{Tianzhe Xu and Joel England\\
SLAC National Accelerator Laboratory}
\date{August 31, 2024}

\begin{document}

\maketitle
\begin{abstract}
Magnetic compressors offer an attractive path toward jitter-free bunch compression in MeV-UED. Compared with RF bunchers, magnetic compressors do not introduce additional timing jitter and can be configured to minimize timing jitter due to upstream energy fluctuation. In this work we discuss the principles of designing magnetic compressors for MeV-UED. Start-to-end simulation of a dogleg compressor is presented. Results show a bunch length of 11~fs and arrival time jitter of 8~fs can be achieved at 100 fC charge with small emittance growth.

\end{abstract}
\section{Introduction}

Ultrafast electron diffraction (UED) has emerged as a powerful technique to probe structural dynamics in materials science and chemistry~\cite{weathersby2015mega,Filippetto2022review}. Achieving ultrafast temporal resolution has always been a frontier in the development of UED with the goal of reaching ten or sub-ten femtosecond resolution.  The overall temporal resolution of an UED instrument is usually described by~\cite{li2009temporal},
\begin{equation}
    \tau _ { \text {res } } = \sqrt { \tau _ { \text {pump } } ^ { 2 } + \tau _ { \text {probe } } ^ { 2 } + \tau _ { \mathrm { VM } } ^ { 2 } + \tau _ { \text {jitter } } ^ { 2 } }
\end{equation}
where $\tau _ { \text {pump } }$ is the duration of pump laser, $\tau _ { \text {probe } }$ is the duration of electron bunch, $\tau _ { \mathrm { VM } }$ is the velocity mismatch, and $\tau _ { \text {jitter } } $ is the fluctuations in temporal delay between laser and electron bunch at sample plane. For mega-electron-volt ultrafast electron diffraction (MeV-UED) systems~\cite{weathersby2015mega}, the velocity mismatch term is negligible and the temporal resolution is usually dominated by $\tau _ { \text {probe } }$ and $\tau _ { \text {jitter } }$.  

When the electron beam is accelerated or bunched by an cavity, fluctuations in cavity phase and amplitude will induce an arrvial time jitter and an energy jitter. The latter translates into a temporal jitter in the drift section downstream,
\begin{equation}
    \tau _ { \text {jitter,drift } } = \frac{L}{c\beta_{0}^{3}\gamma_{0}^{2}} \frac{\Delta{\gamma}}{\gamma_{0}}
\end{equation}
where $L$ is the drift length, $c$ is the speed of light, $\beta_{0}$ and $\gamma_{0}$ are normalized velocity and the Lorentz factor, $\frac{\Delta{\gamma}}{\gamma_{0}}$ is the energy fluctuation. For a beam with 4.2 MeV kinetic energy ($\gamma_{0}$=9.241), an energy fluctuation of 0.05\% induces a timing jitter of 60~fs in a 3~m drift. To achieve better temporal resolution, it is important to minimize both $\tau _ { \text {probe } } $ and $\tau _ { \text {jitter } } $.

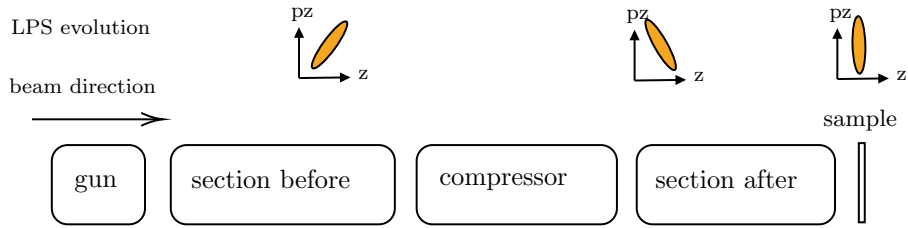
\begin{figure}[t]
    \centering
    \tikzset{every picture/.style={line width=0.75pt}} 

\begin{tikzpicture}[x=0.75pt,y=0.75pt,yscale=-1,xscale=1]

\draw   (54,108) .. controls (54,103.58) and (57.58,100) .. (62,100) -- (93,100) .. controls (97.42,100) and (101,103.58) .. (101,108) -- (101,132) .. controls (101,136.42) and (97.42,140) .. (93,140) -- (62,140) .. controls (57.58,140) and (54,136.42) .. (54,132) -- cycle ;
\draw   (114,108) .. controls (114,103.58) and (117.58,100) .. (122,100) -- (219,100) .. controls (223.42,100) and (227,103.58) .. (227,108) -- (227,132) .. controls (227,136.42) and (223.42,140) .. (219,140) -- (122,140) .. controls (117.58,140) and (114,136.42) .. (114,132) -- cycle ;
\draw   (238,108) .. controls (238,103.58) and (241.58,100) .. (246,100) -- (331,100) .. controls (335.42,100) and (339,103.58) .. (339,108) -- (339,132) .. controls (339,136.42) and (335.42,140) .. (331,140) -- (246,140) .. controls (241.58,140) and (238,136.42) .. (238,132) -- cycle ;
\draw   (349,108) .. controls (349,103.58) and (352.58,100) .. (357,100) -- (441,100) .. controls (445.42,100) and (449,103.58) .. (449,108) -- (449,132) .. controls (449,136.42) and (445.42,140) .. (441,140) -- (357,140) .. controls (352.58,140) and (349,136.42) .. (349,132) -- cycle ;
\draw   (461,99) -- (464,99) -- (464,139) -- (461,139) -- cycle ;
\draw    (44,87) -- (108,87) ;
\draw [shift={(110,87)}, rotate = 180] [color={rgb, 255:red, 0; green, 0; blue, 0 }  ][line width=0.75]    (10.93,-3.29) .. controls (6.95,-1.4) and (3.31,-0.3) .. (0,0) .. controls (3.31,0.3) and (6.95,1.4) .. (10.93,3.29)   ;
\draw  [fill={rgb, 255:red, 245; green, 166; blue, 35 }  ,fill opacity=1 ] (185.52,61.15) .. controls (184.08,60.1) and (186.73,54) .. (191.45,47.53) .. controls (196.17,41.05) and (201.16,36.66) .. (202.61,37.71) .. controls (204.05,38.76) and (201.4,44.86) .. (196.68,51.34) .. controls (191.96,57.81) and (186.96,62.21) .. (185.52,61.15) -- cycle ;
\draw  [fill={rgb, 255:red, 245; green, 166; blue, 35 }  ,fill opacity=1 ] (368.22,62.05) .. controls (366.67,62.93) and (362.2,58) .. (358.25,51.03) .. controls (354.3,44.06) and (352.35,37.7) .. (353.9,36.81) .. controls (355.46,35.93) and (359.92,40.87) .. (363.88,47.83) .. controls (367.83,54.8) and (369.78,61.17) .. (368.22,62.05) -- cycle ;
\draw  [fill={rgb, 255:red, 245; green, 166; blue, 35 }  ,fill opacity=1 ] (461.45,63.93) .. controls (459.66,63.98) and (458.04,57.53) .. (457.83,49.52) .. controls (457.62,41.51) and (458.89,34.98) .. (460.68,34.93) .. controls (462.46,34.88) and (464.08,41.34) .. (464.3,49.35) .. controls (464.51,57.35) and (463.23,63.88) .. (461.45,63.93) -- cycle ;
\draw    (178.67,66.17) -- (201.67,65.87) ;
\draw [shift={(204.67,65.83)}, rotate = 179.27] [fill={rgb, 255:red, 0; green, 0; blue, 0 }  ][line width=0.08]  [draw opacity=0] (5.36,-2.57) -- (0,0) -- (5.36,2.57) -- cycle    ;
\draw    (178.67,66.17) -- (178.67,43.5) ;
\draw [shift={(178.67,40.5)}, rotate = 90] [fill={rgb, 255:red, 0; green, 0; blue, 0 }  ][line width=0.08]  [draw opacity=0] (5.36,-2.57) -- (0,0) -- (5.36,2.57) -- cycle    ;

\draw    (347.92,67.42) -- (370.92,67.12) ;
\draw [shift={(373.92,67.08)}, rotate = 179.27] [fill={rgb, 255:red, 0; green, 0; blue, 0 }  ][line width=0.08]  [draw opacity=0] (5.36,-2.57) -- (0,0) -- (5.36,2.57) -- cycle    ;
\draw    (347.92,67.42) -- (347.92,44.75) ;
\draw [shift={(347.92,41.75)}, rotate = 90] [fill={rgb, 255:red, 0; green, 0; blue, 0 }  ][line width=0.08]  [draw opacity=0] (5.36,-2.57) -- (0,0) -- (5.36,2.57) -- cycle    ;

\draw    (450.17,66.67) -- (473.17,66.37) ;
\draw [shift={(476.17,66.33)}, rotate = 179.27] [fill={rgb, 255:red, 0; green, 0; blue, 0 }  ][line width=0.08]  [draw opacity=0] (5.36,-2.57) -- (0,0) -- (5.36,2.57) -- cycle    ;
\draw    (450.17,66.67) -- (450.17,44) ;
\draw [shift={(450.17,41)}, rotate = 90] [fill={rgb, 255:red, 0; green, 0; blue, 0 }  ][line width=0.08]  [draw opacity=0] (5.36,-2.57) -- (0,0) -- (5.36,2.57) -- cycle    ;

\draw (64,114) node [anchor=north west][inner sep=0.75pt]   [align=left] {gun};
\draw (123,111) node [anchor=north west][inner sep=0.75pt]   [align=left] {section before};
\draw (357,111) node [anchor=north west][inner sep=0.75pt]   [align=left] {section after};
\draw (248,113) node [anchor=north west][inner sep=0.75pt]   [align=left] {compressor};
\draw (441.75,80.75) node [anchor=north west][inner sep=0.75pt]   [align=left] {{\small sample}};
\draw (31,65) node [anchor=north west][inner sep=0.75pt]   [align=left] {{\footnotesize beam direction}};
\draw (32,35) node [anchor=north west][inner sep=0.75pt]   [align=left] {{\footnotesize LPS evolution}};
\draw (207.17,60.42) node [anchor=north west][inner sep=0.75pt]  [font=\scriptsize] [align=left] {{\footnotesize z}};
\draw (173.17,29.25) node [anchor=north west][inner sep=0.75pt]  [font=\scriptsize] [align=left] {{\footnotesize pz}};
\draw (342.42,30.5) node [anchor=north west][inner sep=0.75pt]  [font=\scriptsize] [align=left] {{\footnotesize pz}};
\draw (376.42,61.67) node [anchor=north west][inner sep=0.75pt]  [font=\scriptsize] [align=left] {{\footnotesize z}};
\draw (444.67,29.75) node [anchor=north west][inner sep=0.75pt]  [font=\scriptsize] [align=left] {{\footnotesize pz}};
\draw (478.67,60.92) node [anchor=north west][inner sep=0.75pt]  [font=\scriptsize] [align=left] {{\footnotesize z}};

\end{tikzpicture}
    \caption{Diagram for an UED beamline with a magnetic compressor. The sections before and after include non-bending elements like drift, quadrupoles or solenoids. The evolution of longitudinal phase space (LPS) is shown on top (bunch head on the right).}
    \label{fig:compressor_diagram}
\end{figure}

Various methods have been proposed and demonstrated to improve the temporal resolution such as THz-based compression~\cite{zhao2020,Snively2020} and direct jitter measurements with a timing tool~\cite{zhao2018,zhao2019,li2019,othman2024improved}. Among these methods, isochronous magnetic compressor~\cite{kim_towards_2020,qi2020} is an attractive option. As depicted in Figure~\ref{fig:compressor_diagram}, magnetic compressor can be configured to impart a negative $R_{56}$ and compress eletron bunch duration $\tau _ { \text {probe } }$ while simultaneously canceling the timing jitter $\tau _ { \text {jitter,drift} } $. In this work we discuss the requirements of magnetic compressors for MeV-UED and describe the procedures for obtaining an optimal design. We present the design of a dogleg compressor for the SLAC MeV-UED beamline~\cite{weathersby2015mega} and show a compressed electron beam with 11~fs bunch length can be generated with the current RF gun.

\section{Design requirements procedures}

\begin{figure}[!h]
     \centering
     \begin{subfigure}{0.4\textwidth}
         \includegraphics[width=\textwidth]{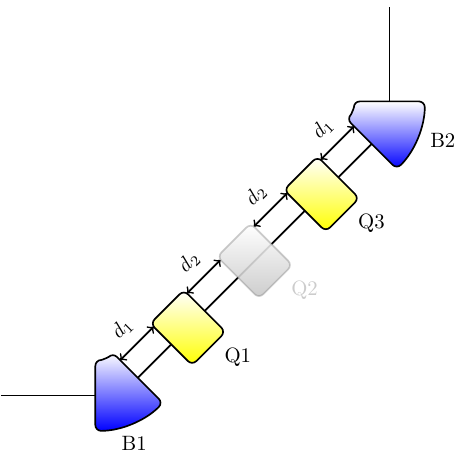}
         \caption{achromat}
         \label{fig:achromat_diagram}
     \end{subfigure}
     \hfill
     \begin{subfigure}{0.5\textwidth}
         \includegraphics[width=\textwidth]{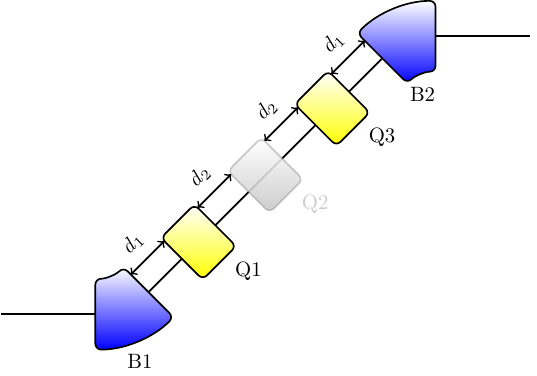}
         \caption{dogleg}
         \label{fig:dogleg_diagram}
     \end{subfigure}
     
     \vspace{4ex}
          \begin{subfigure}{0.8\textwidth}
         \includegraphics[width=\textwidth]{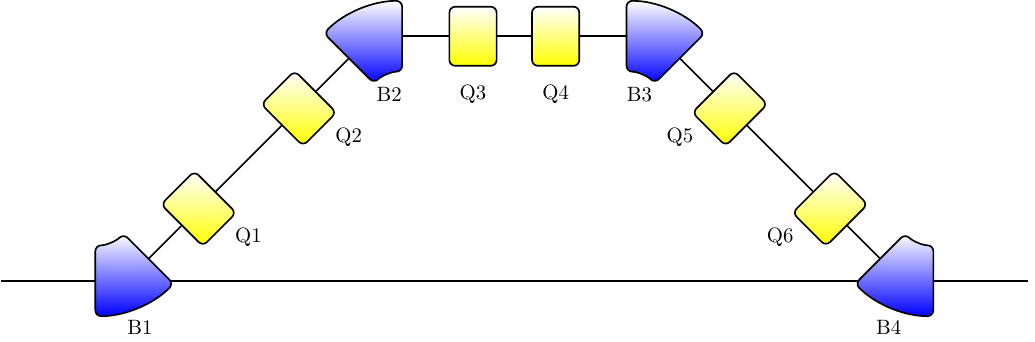}
         \caption{chicane}
         \label{fig:chicane_diagram}
     \end{subfigure}
     
        \caption{Different types of compressors. (a) achromat and (b) dogleg with optional quadrupole magnet Q2 (grayed out) in the center. (c) Chicane.}
        \label{fig:compressor_type}        
\end{figure}

We mainly consider two types of magnetic compressors depicted in Figure~\ref{fig:compressor_type}(a) and (b), the double bend achromat and the dogleg. We will refer to the former as `achromat' for simplicity. Both consist of two bending magnets and three quadrupoles (the middle one is optional). There are four requirements for the lattice:
\begin{enumerate}
    \item Achieves first-order achromat condition, i.e., $\eta = \eta^{\prime}=0$ at compressor exit.
    \item Minimize emittance growth.
    \item Achieves isochronous condition, i.e., $R_{56,\text{total}}=R_{56,\text{before}}+R_{56,\text{compressor}}+R_{56,\text{after}}=0$ for the full beamline. 
    \item Fully compresses electron bunch at sample location.
\end{enumerate}

In the following we will treat requirement 1\&2 (transverse dynamics) and requirements 3\&4 (longitudinal dynamics) separately.

\subsection{Transverse dynamics}\label{sec:transverse}

\begin{figure}[!b]
    \centering
    \includegraphics[width=0.93\linewidth]{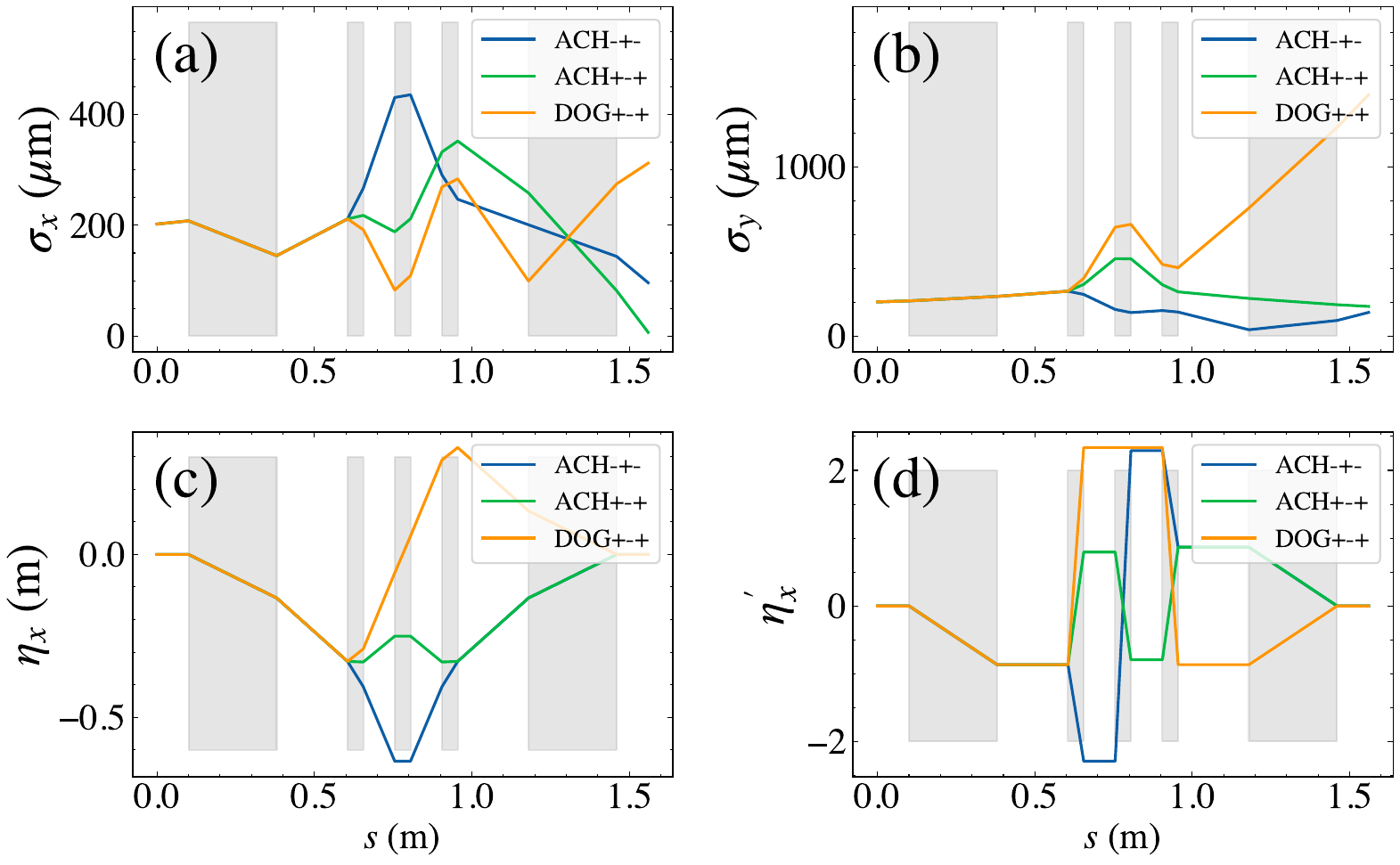}
    \caption{Evolution of beam sizes, dispersions and dispersion slopes for different beamline configurations. The gray bar regions denote dipole and quadrupole magnets.}
    \label{fig:beamlines_evo}
\end{figure}

The first-order achromat condition is achieved by setting the quadrupole magnets to appropriate strengths. Evolution of dispersion $\eta$ in the compressor can be described by,
\begin{equation}
    \eta ^ { \prime \prime } + K ( s ) \eta = \frac{1}{\rho ( s )}
\end{equation}
where $K(s)$ and $\rho(s)$ are the focusing strength and bending radius along the beamline. In addition to the requirements above, we also enforce the symmetry condition of $K(s)$ around midpoint. Since $\rho(s)$ is symmetric (even) for achromat and antisymmetric (odd) for dogleg, and that $\eta(s)$ and $\eta ^ { \prime \prime } (s)$ have the same symmetry, we can deduce $\eta(s)$ is even for achromat and odd for dogleg (see for example in Figure~\ref{fig:beamlines_evo} (c) and (d)). Therefore, at the midpoint of the compressor, $\eta^{\prime}=R_{26}=0$ for achromat, and $\eta=R_{16}=0$ for dogleg. For two-quad configurations, it allows us to solve for the focal strength of the quadrupole under thin lens approximation,
\begin{equation}
    q = \frac { 1 } { f } = \frac { \sin \theta } { d _ { 1 } \sin \theta - \rho ( \cos \theta - 1 ) }
\end{equation}
for achromat. And 
\begin{equation}
    q = \frac { 1 } { f } = \frac { \rho ( \cos \theta - 1 ) - \left( d _ { 1 } + d _ { 2 } \right) \sin \theta } { d _ { 2 } \rho ( \cos \theta - 1 ) - d _ { 1 } d _ { 2 } \sin \theta }
\end{equation}
for dogleg.

Another observation based on the symmetry analysis is the different quad strengths required to match the dispersion condition for the two types of compressors. While the achromat beamline only needs to zero $\eta^{\prime}$ at the midpoint, the dogleg beamline needs to flip the sign of $\eta^{\prime}$ with the quadrupoles. Therefore it requires higher quadrupole strengths than the achromat beamline.

A third quadrupole can be inserted in the middle to control the envelope evolution of the bunch. For achromat and dogleg beamlines, there are generally three quadrupole solutions that fulfill the achromat condition, ``ACH-+-", ``ACH+-+", ``DOG+-+" (``+" denotes focusing in $x$ and ``-" denotes defocusing in $x$). A quadrupole setting of ``DOG-+-" would not exist due to requirement of $\eta=0$ at midpoint. Typical evolutions of beam sizes, dispersions and dispersion slope of the three solutions are shown in Figure~\ref{fig:beamlines_evo}. 

\begin{figure}[h]
    \centering
    \includegraphics[width=1.0\linewidth]{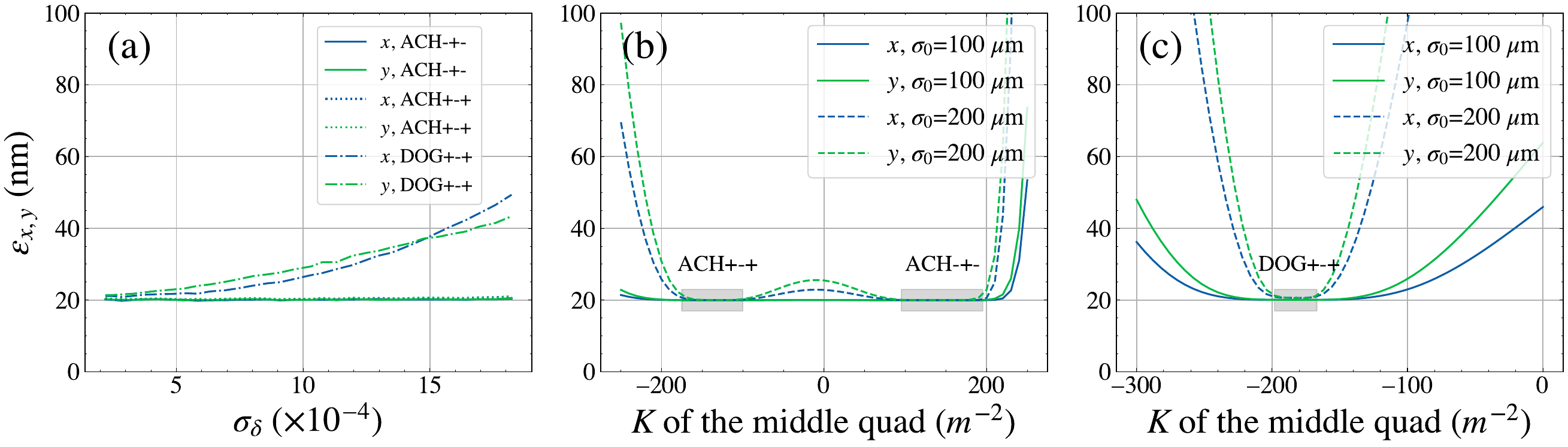}
    \caption{(a) Final emittances for electron beams with different energy spread. (b) Final emittances for electron beams with different initial spot size in an achromat compressor. (c) Final emittances for electron beams with different initial spot size in a dogleg compressor. The gray bars in (b) and (c) denote the ranges where emittance growth is minimized.}
    \label{fig:emit_scan}
\end{figure}

Beyond first-order considerations, second-order effects such as chromatic and geometric aberrations can induce emittance growth through the compressor beamline. The effect of chromatic terms ($T_{**6}$) is usually small as the energy spread of the electron beam in an UED beamline is on the order of $10^{-4}$. Here we consider the emittance growth in the three-quadrupole compressor lattices for a low-emittance (20~nm) electron beam with different initial energy spread. As shown in Figure~\ref{fig:emit_scan}(a), the emittance growth will occur when the rms energy spread increases to $10^{-3}$ level and is more prominent for dogleg as it requires higher quadrupole strengths.  For geometric aberrations, care needs to be taken when matching the electron beam into the compressor lattice to minimize emittance growth. Again we consider the low-emittance electron beam with different initial spot sizes. For the two types of compressor lattice (achromat and dogleg), we vary the strengths of the middle quadrupole Q2 and adjust the strengths of Q1 and Q3 to achieve achromat condition. As shown in Figure~\ref{fig:emit_scan}(b) and (c), for both beamline configurations there exists an optimal range of Q2 strengths where emittance growth is minimized. The range is wider for smaller initial size than for larger initial size.

In addition to achromat and dogleg, a symmetric chicane compressor can also be used for bunch compression and jitter suppression in MeV-UED. Unlike the configurations often used in FEL which expect a negative chirp, the quadrupole strengths need to be adjusted to provide a negative $R_{56}$. As shown in Figure~\ref{fig:compressor_type} (c), a chicane compressor can be regarded as two cascaded doglegs. Since the beam will exit the second dipole with a divergent beam size, matching optics are needed to refocus the beam and minimize emittance growth in the second dogleg. The advantage of the chicane configuration is that it doesn't introduce trajectory offset and the dipole can be turned off to bypass the compressor when needed. The design consideration for the chicane is similar to double bend achromats and doglegs and will not be discussed in this work. 

So far we've only considered the lattice design based on single particle dynamics. In reality, space charge will play a role in transverse dynamics: (1) the defocusing due to space charge alters the beam size and dispersion evolution and the quadrupole strengths need to be retuned to achieve achromat condition, (2) when focused to a small spot size, space charge will induce emittance growth for the electron beam. Both effects are beyond our transfer matrix treatment and require a full scale simulation which will be presented in Section~\ref{sec:s2e-simulation}.

\subsection{Longitudinal dynamics}\label{sec:longitudinal}

\begin{figure}[h]
    \centering
    \includegraphics[width=0.95\linewidth]{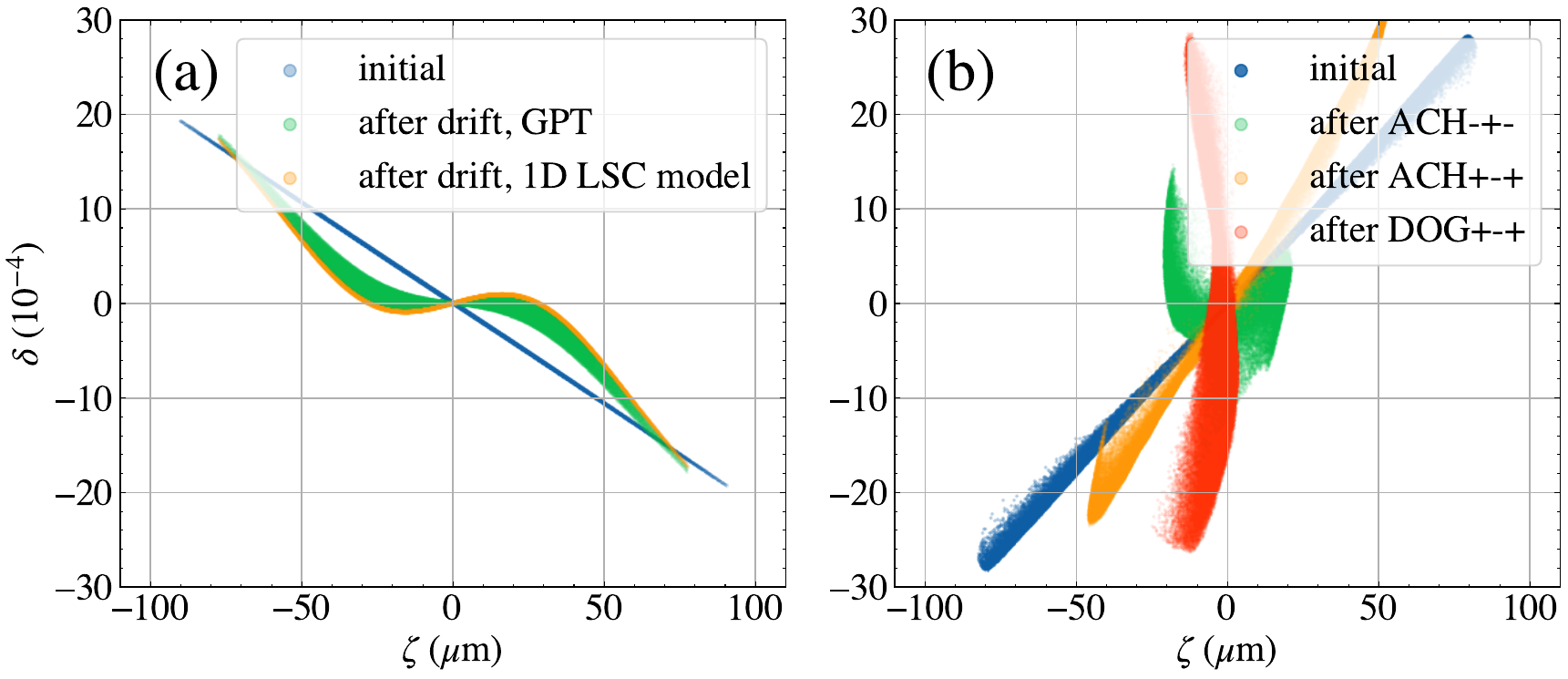}
    \caption{(a) Evolution of LPS after a drift for an artificial beam. (b) Comparison of LPS for different compressor solutions.}
    \label{fig:lps-compare}
\end{figure}

To meet requirement 3 and 4, the bending radius and angle of the compressor needs to be chosen to match the total path length from gun exit to sample. The location of the compressor also needs to be optimized so the beam is fully compressed at sample location for the incoming chirp. To determine the location of longitudinal focus, we needs to incorporate  single particle dynamics and longitudinal space charge (LSC) as the later can influence the evolution of longitudinal phase space (LPS) drastically. For design purposes, we combine the first- and second-order transfer matrix approach with a one-dimensional impedance model~\cite{huang2006,qiang2009high,codeexample} which allows fast modelling of LPS evolution through compressor beamline\footnote{Similar approach is achieved in {\sc elegant} with {\sc lscdrift} element or in {\sc ocelot}. The difference in our implementation is that we use the impedance averaged over transverse density while {\sc elegant} uses the impedance on axis.}. To benchmark the model, we generate an artificial beam distribution with 100 fC charge and track its $(\zeta,\delta)$ LPS evolution in a 0.6~m drift with {\sc gpt} and the 1D model. Here $\zeta$ is the relative longitudinal position within the bunch defined as $\zeta=\langle s \rangle-ct$ and $\delta=\frac{p-p_{0}}{p_{0}}$ is the momentum deviation. The final LPS are compared in Figure~\ref{fig:lps-compare}(a) and the two codes show good agreement. The 1D model fails to model the increase in uncorrelated energy spread as it does not account for radial dependence of space charge forces. However, it is still useful for our compressor design as it predicts the evolution of the chirp which determines the fully-compressed location.

The $R_{56}$ of the three compressor configurations ``ACH-+-", ``ACH+-+", ``DOG+-+" is given by
\begin{equation}
    R_{56,\text{compressor}} = \frac { L _ { \mathrm { mid } } + 2 \rho \theta } { \gamma ^ { 2 } } - 2 \rho ( \theta - \sin \theta )
\end{equation}
where $ L _ { \mathrm { mid } }$ is the total length between two bends. In linear approximation, the three configurations should apply the same transformation to LPS. However, when LSC is included, the LPS evolution for the three compressor configurations can be quite different. Here we consider a positively-chirped beam with 100 fC charge exiting from RF gun with its LPS shown in Figure~\ref{fig:lps-compare}(b). The beam is tracked through three different compressor lattices and reaches nominal compression only in the ``DOG+-+" configuration. The differences in the final LPS come from the different evolution of transverse beam sizes in the compressor lattices. For the two achromat configurations, the beam sizes stay focused throughout the compressor; see Figure~\ref{fig:beamlines_evo}(a) and (b). As a result the LSC forces are stronger than the dogleg configurations and induce nonlinear distortion in LPS and counteract the compression when the chirp is flipped. 

\subsection{Procedures}\label{sec:procedures}

We'll end this section with a summary of our procedures in reaching a compressor design. Given the nominal operation settings of gun and laser, we obtain the chirp at the gun exit and the $R_{56}$ required to fully compress the electron bunch. The type and layout of the compressor beamline (drift lengths, bending radii and angles) can then be determined based on the $R_{56}$, the isochronous condition, and total space available. Quadrupole solutions of the compressor are determined from the transverse emittance growth as well as beam size evolution. The 1D LSC code is used to confirm whether the beam reaches longitudinal focus at sample location by iterating between different beamline layouts, quadrupole settings, and solenoid settings upstream to vary the initial beam size and chirp entering the compressor. Once it's confirmed in the matrix code that requirements 1-4  are satisfied, the full beamline is simulated and optimized in a full scale PIC code {\sc gpt}~\cite{gpt}. The result of an optimized beamline is presented in the following section.

\section{Start-to-end simulation}\label{sec:s2e-simulation}

\begin{figure}[b]
    \centering
    \includegraphics[width=0.95\linewidth]{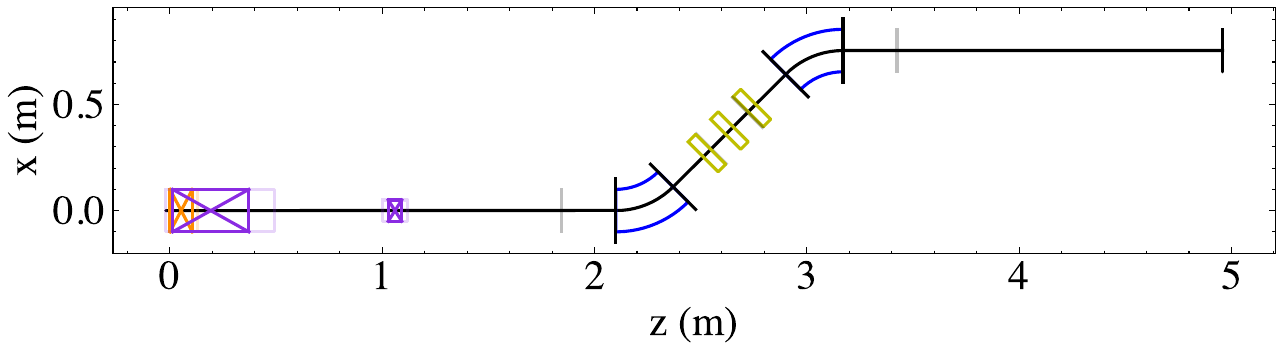}
    \caption{Layout of a dogleg compressor beamline for SLAC MeV-UED. The orange,  purple, yellow and blue boxes denote gun, solenoids, quadrupoles and dipoles. The vertical bar at z=5m is the sample location.}
    \label{fig:dogleg_beamline}
\end{figure}

\begin{table}[h]
    \centering
    \begin{tabular}{ccc} \hline\hline
    

           \multicolumn{3}{c}{compressor beamline} \\ \hline
        $\rho$ & 0.382 & m\\ 
        $\theta$ & 45 & deg\\ 
        $d_{1}$ & 0.2 & m\\ 
        $d_{2}$ & 0.1 & m\\ 
        $l_{q}$ & 0.05 & m\\ 
        Gradient of quadrupole $Q{1}$ & -3.30 & T/m\\ 
        Gradient of quadrupole $Q{2}$ & 3.15 & T/m\\ 
        Gradient of quadrupole $Q{3}$ & -3.29 & T/m\\ 
        Distance from gun exit to compressor entrance & 1.97 & m\\ 
        Distance from compressor exit to sample plane & 1.7875 & m\\ 
         \hline
         \multicolumn{3}{c}{electron beam} \\ \hline
        Charge & 100 & fC\\ 
        Kinetic energy & 4.2 & MeV\\ 
        \hline
        $\sigma_{t}$ before compressor & 112.8 & fs\\ 
        $\sigma_{\text{jitter}}$ before compressor, TS1 & 36 & fs\\ 
        $\sigma_{\text{jitter}}$ before compressor, TS8 & 8 & fs\\ 
        $\varepsilon_{x,y}$ before compressor & 26.8,~26.8 & nm\\  
        \hline
        $\sigma_{t}$ after compressor & 11.6 & fs\\         
        $\sigma_{\text{jitter}}$ after compressor, TS1 & 21 & fs\\         
        $\sigma_{\text{jitter}}$ after compressor, TS8 & 8 & fs\\         
        $\varepsilon_{x,y}$ after compressor & 29.3,~27.7 & nm\\ 
         \hline\hline
    \end{tabular}
    \caption{Parameters for the compressor beamline and electron beam parameters before ($z=1.8$m) and after compressor (at sample).}
    \label{tab:parameters}
\end{table}

Here we present the start-to-end simulation of a dogleg compressor for the SLAC MeV-UED beamline. The simulation is performed with {\sc gpt} and includes 3D space charge effects and realistic fringe fields of the magnets. The layout of the beamline is illustrated in Figure~\ref{fig:dogleg_beamline}. The parameters of the compressor and the electron beam are summarized in Table~\ref{tab:parameters}. The beam line consists of a 1.6 cell S-band RF gun and two solenoids for matching the electron beam into the compressor. The strength of the solenoid is chosen to minimize the geometric aberration in the dogleg as discussed in section~\ref{sec:transverse}. The beam then undergoes compression in the dogleg and a final rms bunch length of 11.6 fs is achieved with small emittance growth (2.5~nm in $x$ and 0.9~nm in $y$).

\begin{figure}[h]
    \centering
    \includegraphics[width=0.98\linewidth]{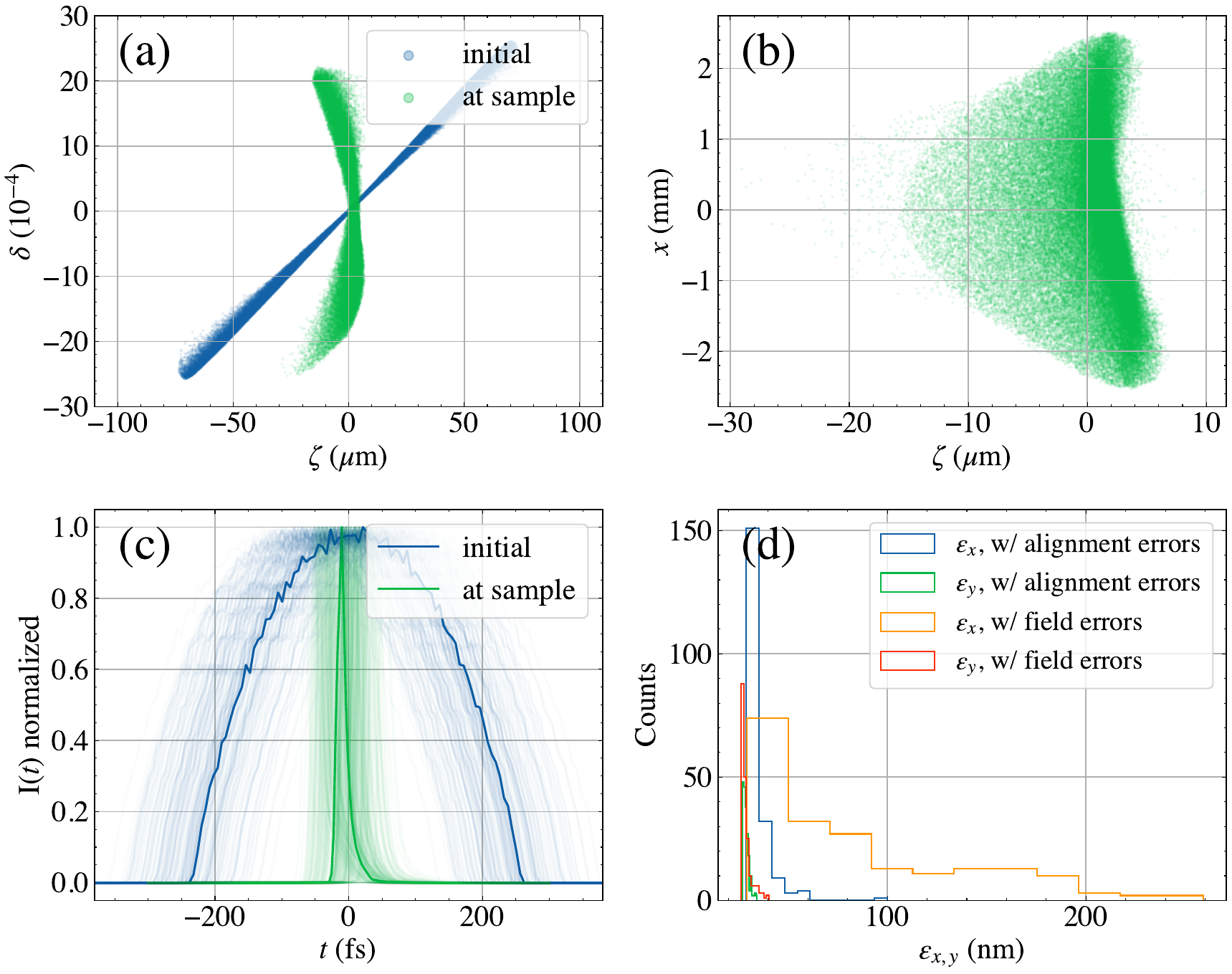}
    \caption{(a) Longitudina phase space before and after compression. (b) ($\zeta$,$x$) phase space at sample. (c) Temporal profiles of the bunch before and after compression. The transparent and solid traces are profiles with and without RF jitters. (d) Histogram of final emittances due to alignment and magnet field errors }
    \label{fig:dogleg_s2e}
\end{figure}

The LPS before and after compression are shown in Figure~\ref{fig:dogleg_s2e}(a). We can see the chirp is removed and beam is fully compressed at sample location. Due to second-order effects of the compressor, the beam has developed some nonlinear correlation in the longitudinal and spatiotemporal phase space;see Figure~\ref{fig:dogleg_s2e}(a) and (b). As a result, collimation at beam center after the compressor only reduces the emittance of the beam but not its bunch length because the center slice has the longest bunch length.

To confirm that the time-of-flight jitters $\tau _ { \text {jitter,drift } }$ are indeed suppressed by the compressor, we performed 200 start-to-end simulations with different random realizations of the rf amplitude and phase for the gun. The amplitude and phase values were randomly generated with a normal distribution with respective rms values of 50 ppm and 0.05 degree based on klystron performance of at Test Stand 1 (TS1). The temporal profiles of the electron bunch before and after compression are shown in Figure~\ref{fig:dogleg_s2e}(c). Both the temporal length and jitter are reduced with the values summarized in Table~\ref{tab:parameters}. We note that the compressor is only able to compress the timing jitter due to upstream energy fluctuation---the timing jitter due to different time-of-flight in the gun remains. Recently the SLAC MeV-UED beamline has switched to a ScandiNova solid-state klystron at Test Stand 8 (TS8) for operation and  achieved better RF stability with klystron voltage jitter of 20 ppm. The time-of-flight jitters are re-evaluated for the TS8 scenario and the rms jitter value at the sample is found to be 8~fs.

We performed similar randomized simulations to evaluate the sensitivity of the transverse emittances to fluctuations of magnet field strength and quadrupole alignment errors. The rms value of the random normal distribution is set to be 1\% for magnet field strength fluctuations and 200$\mu$m for quadrupole alignment errors in $x$ and $y$. The final transverse emittances are shown in Figure~\ref{fig:dogleg_s2e}(d). While the final vertical emittances are not susceptible to the errors, the growth in horizontal emittances can be quite large especially when magnet field errors are present. 

\section{Conclusion}
In conclusion, we have described the design requirements and procedures of magnetic compressors for MeV-UED. The transverse and longitudinal dynamics in double bend achromat and dogleg are discussed and compared. A code based on transfer matrix and 1D LSC model was developed to model emittance growth and chirp evolution in the compressor and allows fast optimization of compressor parameters. We also present the design of a dogleg compressor for the SLAC MeV-UED beamline. Simulation results show that a 100~fC electron beam with a rms duration of 11~fs and arrival time jitter of 8~fs can be generated and the timing jitter due to upstream energy fluctuation can be minimized.

\bibliographystyle{ieeetr}
\bibliography{sample.bib}
\end{document}